\documentclass[12pt]{article}
\usepackage{amssymb,amsmath,epsfig}
\allowdisplaybreaks
\begin{document}

\title{\bf Impact of Charge on Gravastars in $f(\mathfrak{R},\mathcal{T}^{2})$ Gravity}
\author{M. Sharif \thanks{msharif.math@pu.edu.pk} and Saba Naz
\thanks{sabanaz1.math@gmail.com}\\
Department of Mathematics, University of the Punjab,\\
Quaid-e-Azam Campus, Lahore-54590, Pakistan.}
\date{}
\maketitle

\begin{abstract}
This paper studies the influence of charge on a compact stellar
structure also regarded as vacuum condensate star in the background
of $f(\mathfrak{R},\mathcal{T}^{2})$ gravity. This object is
considered the alternate of black hole whose structure involves
three distinct regions, i.e., interior, exterior and thin-shell. We
analyze these domains of a gravastar for a particular model of this
modified theory. In the inner region of a gravastar, the considered
equation of state defines that energy density is equal to negative
pressure which is the cause of repulsive force on the spherical
shell. In the intermediate shell, pressure and energy density are
equal and contain ultra-relativistic fluid. The inward-directed
gravitational pull of thin-shell counterbalance the force exerted by
the inner region of a gravastar allowing the formation of a
singularity-free object. The Reissner-Nordstrom metric presents the
outer vacuum spherical domain. Moreover, we discuss the impact of
the charge on physical attributes of a gravastar such as the
equation of state parameter, entropy, proper length and energy. We
conclude that singularity-free solutions of charged gravastar are
physically consistent in this modified theory.
\end{abstract}

\section{Introduction}

Cosmic evolution is an accumulation of substantial changes including
dark energy and gravitational collapse, which has developed the
interest of researchers to debate on a variety of problems of
cosmology as well as gravitational physics. When the fuel in the
core of a star runs out and there is not enough pressure to
counteract the strong gravitational pull, the star collapses,
resulting in the formation of new compact objects. The rapid
expansion of the cosmos is confirmed through several astrophysical
observations such as supernova type Ia and cosmic microwave
background. Alternative gravitational theories support expanding
behavior of the universe. These theories are considered the most
favorable and remarkable techniques to discover hidden attributes of
the cosmos. These approaches are constructed by adding higher order
curvature invariants in the geometric part of the Einstein-Hilbert
action. The $f(\mathfrak{R})$ gravity is the natural extension of
general relativity (GR) which discusses some novel aspects of cosmos
\cite{3a}.

Many researchers have shown a keen interest in the idea of curvature
and matter coupling. These modified proposals explain the rotation
curves of galaxies and various cosmic eras. The conservation law
does not hold in these theories, assuring the existence of an extra
force on particles. These theories are assumed to be an excellent
candidate to understand the dark cosmos. The minimal
curvature-matter coupling was developed in \cite{4a} named as
$f(\mathfrak{R},T)$ gravity, while non-minimal coupled theory is
$f(\mathfrak{R}, {T},{\mathfrak{R}}_{\gamma\delta}
{T}^{\gamma\delta})$ gravity \cite{5a}. The presence of
singularities in GR is a major problem due to their prediction in
high energy regions, where GR is not viable due to quantum effects
and there is no specific approach in quantum theory. Accordingly, a
new modification of GR has been developed by adding the analytic
function ${T}_{\gamma\delta}{T}^{\gamma\delta}=\mathcal{T}^{2}$ in
the functional action of $f(\mathfrak{R})$ theory \cite{7a}, dubbed
as $f(\mathfrak{R},\mathcal{T}^{2})$ gravity. This modified theory
is also known as energy-momentum squared gravity (EMSG) which
includes squared components of the matter variables, that can be
used to examine several interesting astrophysical consequences. In
the early universe, this theory has a regular bounce, i.e., finite
maximum energy density as well as small-scale parameter \cite{8a}.
As a result, the big-bang singularity can be resolved using this
non-quantum prescription.

It is worth noting that this gravity overcomes the singularity of
spacetime without affecting the cosmological evolution. Board and
Barrow \cite{9a} investigated exact solutions with perfect matter
configuration in the presence of a specific EMSG model. Nari and
Roshan \cite{10a} studied dense objects that were stable as well as
physically realistic. Bahamonde et al. \cite{11a} explored different
EMSG coupling models (\textit{minimal}/\textit{non-minimal}) and
found that these models unveil mysteries of the cosmos. This theory
incorporates significant work and illustrates numerous cosmic
implications \cite{12a}.

A hypothetical highly compact star free from a singularity is
suggested as a promising substitute for the black hole, which may be
developed by scrutinizing the essential concept of Bose-Einstein
condensation, known as gravastar. Mazur and Motolla \cite{13a}
proposed this unique model which provides the corresponding solution
of the Einstein field equations, they fabricated this cold and dark
object as a gravitationally vacuum star or gravastar. Several
researchers are interested to study the structure of gravastar
because this model is expected to resolve two major issues
associated with black hole structure: the singularity formation and
the information paradox. The gravastar structure comprises of three
different domains, interior region, thin-shell, and exterior region.
The connection between state variables in the inner domain is of the
type $\varrho=-\mathcal{P}$, where $\varrho$ and $\mathcal{P}$
represent energy density and pressure, respectively. These matter
configuration yields non-attractive force which is responsible to
produce enough pressure to resist the collapsing phenomenon. This
interior zone is guarded by a thin-shell that holds stiff fluid
obeying the following relationship $\varrho=\mathcal{P}$. The inner
domain experiences inward-directed force exerted by the shell,
consequently hydrostatic equilibrium is maintained and a
singularity-free compact object is formed. The exterior domain is
completely vacuumed and satisfies the equation of state (EoS)
$\varrho=0=\mathcal{P}$.

There is no observational evidence in the support of gravastar, but
few indirect evidence justifies the existence of gravastar
structure. Sakai et al. \cite{d1} provided the criteria to detect
the gravastar by investigating their shadows. Kubo and Sakai
\cite{d2} used gravitational lensing to detect gravastar by
observing maximum luminosity effects. But, these effects are not
found in black holes with equal mass. Moreover, using
interferometric LIGO detectors, a cosmic event GW150914 observed a
ringdown signal \cite{d3}-\cite{d4}. These signals are released by
objects having no event horizon, thus, strongly suggesting the
presence of gravastar. Recently, an image taken by the first M87
Event Horizon Telescope has been examined and found likely to
gravastar \cite{d5}.

Visser and Wiltshire \cite{5aa} investigated the stability of
gravastars in GR corresponding to specific EoS and found dynamically
stable gravastar structures. Carter \cite{6aa} demonstrated
non-singular solutions of gravastars to examine their various
characteristics. Horvat and Iliji\'{c} \cite{7ddd} studied the
gravastar configuration by taking inner de Sitter and outer
Schwarzschild black hole. There is extensive literature on gravastar
solutions with different matter contents \cite{9aa}-\cite{14aa}.
Ghosh et al. \cite{cd} analyzed higher-dimensional gravastars and
results were compared with 4-dimensional analog model. The
astonishing findings of gravastars inspired the research community
to analyze their physical attributes in modified theories
\cite{29aa}-\cite{65aa}.

Electromagnetism plays a crucial role in the study of structure
evolution and stability of collapsing celestial objects. To maintain
the equilibrium state of a stellar object, a star needs an enormous
amount of charge to overcome the strength of the gravitational pull.
Lobo and Arellano \cite{cg1} developed gravastar solutions
incorporating nonlinear electrodynamics and investigated its
significant structural properties. Horvat et al. \cite{cg2}
discussed charged gravastar and computed surface redshift, the EoS
parameter as well as the speed of sound for the model under
consideration. Turimov et al. \cite{cg3} provided a brief study on
slowly rotating gravastars by taking into account extremely
magnetized perfect matter. Usmani et al. \cite{cg4} used conformal
motion to investigate gravastar by taking charged interior and
Reissner-Nordstrom metric in the exterior. It may be noted that
charged interior behaves as an electromagnetic mass model and plays
a key role in the stability of gravastar structure by generating
gravitational mass. Rahaman et al. \cite{cg5} explored gravastar
structure in 3-dimensional spacetime with the contribution of charge
and investigated several viable features for such compact
structures. Ghosh et al. \cite{cg6} established a charged gravastar
model in a higher-dimensional manifold and found that physical
attributes ensure the viability of the model. Sharif and Javed
\cite{7dddd} investigated the stability of gravastars in the
framework of quintessence as well as regular black hole geometries
and obtained a continuously increasing profile of physical
characteristics corresponding to thickness of thin-shell.

Sharif and Waseem \cite{cg8} analyzed charged gravastar structure
using conformal motion in realm of $f(\mathfrak{R}, T)$ gravity.
Yousaf et al.  \cite{cg7} studied charged gravastar in the same
gravity and discussed its stability. Bhatti et al. \cite{w2}
explored the stability of charged gravastar in the modified gravity.
Bhar and Rej \cite{w3} studied gravastar admitting conformal motion
to analyze the contribution of charge on the stability of the
considered model. Bhatti and his collaborators \cite{w4} also
investigated gravastars in the realm of ${f(\mathcal{G})}$ gravity,
where $\mathcal{G}$ defines Gauss-Bonnet invariant. They
investigated different attributes of gravastar with and without
electromagnetic field. Various attributes of gravastar solutions are
also analyzed through the gravitational decoupling technique
\cite{w5}. We have studied gravastar structure with Kuchowicz metric
in the context of EMSG gravity \cite{w6} . Recently, Sharif and
Saeed \cite{w7} studied charge-free gravastar accepting conformal
motion in the background of $f(\mathfrak{R},\mathcal{T}^{2})$
gravity and discussed the behavior of various physical attributes.
Motivated by numerous works presenting effects of charge, we are
interested to analyze the influence of charge on the gravastar model
in $f(\mathfrak{R},\mathcal{T}^2)$ gravity.

This paper explores the influence of charge on the geometry of
gravastar in the background of $f(\mathfrak{R},\mathcal{T}^{2})$
gravity. We analyze its structure for the specific EMSG model using
three different domains and also discuss physical attributes
graphically. The paper is organized in the following format. Section
\textbf{2} demonstrates the basic formalism of this theory under the
impact of charge. Section \textbf{3} investigates the structure of
charged gravastar with relevant EoS for each region and section
\textbf{4} reflects various physical features of charged gravastar
including EoS parameter, entropy, proper length as well as energy.
We summarize the results in the last section.

\section{Energy-momentum Squared Gravity}

The action of EMSG theory with the inclusion of charge is classified
as \cite{7a}
\begin{equation}\label{1}
\mathcal{I}=\frac{1}{2\kappa^2}\int
d^4xf\left(\mathfrak{R},{\mathcal{T}}^{2} \right){\sqrt{-g}}+\int
d^4x (\mathcal{L}_{m}+\mathcal{L}_{e})\sqrt{-g},
\end{equation}
where $g$ exhibits determinant of the metric tensor and
$\kappa^{2}=8\pi$ is the coupling constant. Moreover,
$\mathcal{L}_{e}=\frac{-1}{16\pi}\mathcal{F}^{\gamma\delta}\mathcal{F}_{\gamma\delta}$
where
$\mathcal{F}_{\gamma\delta}=\varphi_{\delta,\gamma}-\varphi_{\gamma,\delta}$
represents the electromagnetic field tensor and $\varphi_{\gamma}$
demonstrates the four potential. In comparison to GR, the action
includes additional degrees of freedom which enhance the chances to
get analytical solutions. Due to the existence of an extra force and
matter-dominated era, this theory offers a platform to gain
significant outcomes. The variation of the action with respect to
the metric tensor provides the equations of motion
\begin{equation}\label{2}
\mathfrak{R}_{\gamma\delta}f_{\mathfrak{R}}-\frac{1}{2}g_{\gamma\delta}f+g_{\gamma\delta}\Box
f_{\mathfrak{R}}-\nabla _{\gamma}\nabla_{\delta}f_{\mathfrak{R}}
=\kappa^{2}T_{\gamma\delta}-\Theta_{\gamma\delta}f_{\mathcal{T}^{2}}+\kappa^{2}\mathrm{E}_{\gamma\delta},
\end{equation}
where $f\equiv f(\mathfrak{R}, \mathcal{T}^{2})$, $\Box=
\nabla_{\gamma}\nabla^{\gamma}$, $f_{\mathfrak{R}}= \frac{\partial
f} {\partial \mathfrak{R}}$, $f_{\mathcal{T}^{2}}= \frac{\partial f}
{\partial\mathcal{T}^{2}}$,
\begin{eqnarray}\label{3}
\Theta_{\gamma\delta}
=-2\mathcal{L}_{m}\left(T_{\gamma\delta}-\frac{1}{2}g _{\gamma\delta
}T\right)-4\frac{\partial^{2}\mathcal{L}_{m}}{\partial
g^{\gamma\delta}
\partial g^{\theta\eta}}T^{\theta\eta}-TT_{\gamma
\delta}+2T_{\gamma}^{\theta}T_{\delta\theta},
\end{eqnarray}
and $\mathrm{E}_{\gamma\delta}$ is the electromagnetic
energy-momentum tensor given by
\begin{equation}\label{}
\mathrm{E}_{\gamma\delta}=
\frac{1}{4\pi}\left(\mathcal{F}^{\theta}_{\gamma}\mathcal{F}_{\delta\theta}-\frac{1}{4}g_{\gamma\delta}
\mathcal{F}^{\theta\eta}\mathcal{F}_{\theta\eta}\right).
\end{equation}

The covariant divergence of Eq.(\ref{2}) turns out to be
\begin{equation}\label{4}
\nabla^{\gamma}T_{\gamma\delta}=\frac{1}{2\kappa^{2}} \left[-f_{
\mathcal{T}^{2}}g_{\gamma\delta}{\nabla^{\gamma}}{\mathcal{T}^{2}}+2{\nabla^{\gamma}}(f_{
\mathcal{T}^{2}}{\Theta_{\gamma\delta}})+\nabla^{\gamma}\mathrm{E}_{\gamma\delta}\right].
\end{equation}
Due to the coupling of matter and geometry, the conservation law is
failed in this theory. The field equations of EMSG reduce to
$f(\mathfrak{R})$ and GR when  $f(\mathfrak{R},\mathcal{T}^{2})=
f(\mathfrak{R})$ and $f(\mathfrak{R},\mathcal{T}^{2})=\mathfrak{R}$,
respectively. In the investigation of different astrophysical and
cosmological scenarios, the distribution of matter is crucial. To
examine features of the charged gravastar, we use isotropic matter
distribution
\begin{equation}\label{5}
T_{\gamma\delta}=\left(\mathcal{P}+\varrho\right)\emph{U}_{\gamma}\emph{U}_{\delta}-g_{\gamma\delta}\mathcal{P},
\end{equation}
where $\emph{U}_{\gamma}$ depicts four-velocity. The matter
distribution has two possible choices of $\mathcal{L}_{m}$, either
we can take it as $-\varrho$, or the other choice is
$\mathcal{L}_{m}=\mathcal{P}$. For minimally coupled theories of
gravity, the distribution is unaffected by the choice of matter
lagrangian \cite{pp}. Here, we take $\mathcal{L}_{m}=\mathcal{P}$
and using Eqs.(\ref{3}) and (\ref{5}), it follows that
\begin{eqnarray}\nonumber
\Theta_{\gamma\delta} = -\left(3\mathcal{P}^2+\varrho^2+4
\mathcal{P}\varrho\right)\emph{U}_{\gamma}\emph{U} _{\delta}.
\end{eqnarray}
Manipulating Eq.(\ref{2}), we have
\begin{equation}\label{6}
\mathbb{G}_{\gamma\delta}=
\frac{1}{f_{\mathfrak{R}}}\left(T_{\gamma\delta}^{c}+\kappa^{2}T_{\gamma\delta}
+\kappa^{2}\mathrm{E}_{\gamma\delta}\right)=T_{\gamma\delta}^{eff},
\end{equation}
where $\mathbb{G}_{\gamma\delta}=\mathfrak{R}_{\gamma\delta}
-\frac{1}{2}\mathfrak{R}g_{\gamma\delta}$ is the Einstein tensor,
$T_{\gamma\delta}^{c}$ denotes the additional impact of EMSG and
$T_{\gamma\delta}^{eff}$ expresses the effective stress-energy given
as
\begin{equation}\label{7}
T_{\gamma\delta}^{eff}=\frac{1}{f_{\mathfrak{R}}}\bigg\{\kappa^{2}T_{\gamma\delta}-g_{\gamma\delta}\Box
f_{\mathfrak{R}}+\nabla_{\gamma}\nabla_{\delta}
f_{\mathfrak{R}}-\Theta_{\gamma\delta}f_{\mathcal{T}^{2}}+
\frac{1}{2}g_{\gamma\delta}\left(f-\mathfrak{R}f_{\mathfrak{R}}\right)\bigg\}.
\end{equation}

The field equations become highly non-linear when derivatives of
multivariate functions are involved and obtaining exact solutions is
more challenging. The solution to non-linear equations is obtained
by employing a particular $f(\mathfrak{R},\mathcal{T}^{2})$ linear
model given as \cite{7a}
\begin{equation}\label{8}
f(\mathfrak{R},\mathcal{T}^{2})=\mathfrak{R}+\zeta \mathcal{T}^{2},
\end{equation}
where $\mathcal{T}^{2}=3\mathcal{P}^{2}+\varrho^{2}$ and $\zeta$ is
an arbitrary model parameter. The inclusion of $\mathcal{T}^{2}$ in
this modified gravity produces more wider form of GR than the
$f(\mathfrak{R})$ and $f(\mathfrak{R},T)$ theories of gravity. The
functional form (\ref{8}) describes the $\Lambda$CDM model and has
extensively been used to solve many cosmological issues \cite{ab}.
It depicts three major epochs of the universe: radiation-dominated,
matter-dominated and de Sitter-dominated, while the obtained
solutions indicate rapid expansion. The corresponding field
equations are
\begin{equation}\label{9}
\mathbb{G}_{\gamma\delta}=8\pi T_{\gamma\delta}+\frac{1}{2}\zeta
g_{\gamma\delta} \mathcal{T}^{2}-\zeta f_{
\mathcal{T}^{2}}{\Theta_{\gamma\delta}}+8\pi\mathrm{E}_{\gamma\delta}.
\end{equation}
The results of GR are retrieved for $\zeta=0$.

To study the interior region of charged gravastar, we choose static
spherical spacetime as
\begin{equation}\label{10}
(ds)^{2}_{-}=e^{\chi(r)}dt^{2}-e^{\xi(r)}dr^{2}-r^{2}d\theta^{2}-r^{2}\sin^{2}\theta
d\phi^{2}.
\end{equation}
The corresponding non-zero Einstein tensor components are given as
\begin{eqnarray}\label{11}
\mathbb{G}_{0}^{0}&=&\Pi(r)\left(e^{\xi}-1+{\xi'r}\right),
\\\label{12}
\mathbb{G}_{1}^{1}&=&\Pi(r)\left(e^{\xi}-{\chi'r}-1\right),
\\\label{13}
\mathbb{G}_{2}^{2} &=&\Pi(r) \left
(\frac{2r}{4}({-{\chi'}+\xi'})+({\chi'\xi'}-2{\chi''}-{\chi'^{2}})\frac{{r}^{2}}{4}\right),
\end{eqnarray}
where $\Pi(r)= e^{-\xi}r^{-2}$ and prime symbolizes radial
derivative. The modified equations are obtained by inserting
Eqs.(\ref{5}) and (\ref{11})-(\ref{13}) in (\ref{9}) as
\begin{eqnarray}\label{14}
&&{\xi'r}-1+e^{\xi}
=\frac{r^{2}}{e^{-\xi}}\big[8\pi\varrho+\frac{\zeta}{2}
(3{\varrho^{2}}+9{\mathcal{P}^{2}}+8\mathcal{P}\varrho)+\frac{{q}^{2}}{{r}^{4}}\big],
\\\label{15}
&&e^{\xi}-{\chi'r}-1=\frac{r^{2}}{e^{-\xi}}\big[-8\pi
\mathcal{P}+\frac{\zeta}{2}({\varrho^{2}}+3{\mathcal{P}^{2}})-\frac{{q}^{2}}{{r}^{4}}\big],
\\\nonumber
&&\frac{r}{2}({\xi'}-{\chi'})-(2{\chi''}+{\chi'^{2}}-{\chi'\xi'})
\frac{r^{2}}{4}=\frac{r^{2}}{e^{-\xi}}\big[\frac{\zeta}{2}({\varrho^{2}}+3{\mathcal{P}^{2}})-8\pi
\mathcal{P}+\frac{{q}^{2}}{{r}^{4}}\big],\\\label{16}
\end{eqnarray}
where $q$ indicates the charge of the interior sphere defined by
\begin{equation}\label{}
q(r)=4\pi\int_{0}^{r}r^{2}\sigma(r)e^{\chi/2}dr,\quad
E(r)=\frac{r^{-2}q(r)}{4\pi }.
\end{equation}
Here, $E(r)$ and $\sigma$ represent the electric field intensity and
surface charge density, respectively.

The non-conserved equation (\ref{4}) provides
\begin{equation}\label{17}
\frac{d\mathcal{P}}{dr}+\chi'(\frac{\varrho+\mathcal{P}}{2})+\aleph^\star-\frac{qq'}{4\pi
r^{4}}=0,
\end{equation}
where $\aleph^\star$ shows the significant contribution of EMSG as
follows
\begin{equation}\nonumber
\aleph^\star=\frac{\zeta}{2}\left(3\mathcal{P}^2+\varrho^2+4
\mathcal{P}\varrho\right)\chi'+\zeta(\varrho\varrho'+3\mathcal{P}\mathcal{P}').
\end{equation}
Using Eq.(\ref{14}), we obtain
\begin{equation}\label{18}
e^{-\xi}=\frac{r-2m}{r}-\frac{\zeta}{3}\Big(\frac{3{\varrho^{2}}
+9{\mathcal{P}^{2}}+8\mathcal{P}\varrho}{2}\Big)r^{2}-\frac{1}{r}
\int\frac{q^{2}}{{r}^{2}}dr,
\end{equation}
where $m=\int4\pi {\varrho}r^{2}dr$ is the gravitational mass and
$\mathcal{H}(r)=\frac{1}{r} \int\frac{q^{2}}{{r}^{2}}dr$ depicts
effect of charge. This provides a relationship between matter
variables and the radial metric coefficient of the considered
spacetime. The hydrostatic equilibrium equation (\ref{17}) reads as
follows
\begin{equation}\nonumber
\frac{d\mathcal{P}}{dr}=\frac{-\frac{\chi'}{2}(\varrho+\mathcal{P})-\zeta
\frac{\chi'}{2}\left(3\mathcal{P}^2+\varrho^2+4
\mathcal{P}\varrho\right)-\zeta\varrho\varrho'++\frac{qq'}{4\pi
r^{4}}}{3\mathcal{P}\zeta+1},
\end{equation}
where $\chi'$ is calculated from Eqs.(\ref{15}) and (\ref{18}) reads
as
\begin{equation}\nonumber
\chi'=\frac{r\big(8\pi
\mathcal{P}-\frac{\zeta}{2}(\varrho^{2}+3{\mathcal{P}^{2}}+\frac{q^{2}}{r^{4}})\big)
+{\frac{1}{r}\big(\frac{2m}{r}+\frac{\zeta}{2}(3{\varrho^{2}}+
9{\mathcal{P}^{2}}+8\mathcal{P}\varrho+\mathcal{H}(r))\big)\frac{r^{2}}{3}}}{1-\frac{2m}{r}
-\frac{\zeta}{2}(3{\varrho^{2}}+9{\mathcal{P}^{2}}+8\mathcal{P}\varrho)-\mathcal{H}(r)}.
\end{equation}
This reduces to the Tolman-Oppenheimer-Volkoff equation of GR for
$\zeta=0$.

\section{Structure of Charged Gravastars}

In this section, we investigate charged gravastar geometry by
considering three regions (inner, thin-shell and outer) satisfying
particular EoS. The shell of negligible thickness covers internal
domain of the gravastar lying in the range
$\mathbf{R}_{2}=\mathbf{R}+\epsilon>r>\mathbf{R}=\mathbf{R}_{1}$,
where negligible thickness of the shell is indicated by
$\mathbf{R}_{2}-\mathbf{R}_{1}=\epsilon$. The radii of the inner as
well as outer regions of charged gravastars are represented by
$\mathbf{R}_{1}$ and $\mathbf{R}_{2}$, respectively. The three
regions of charged gravastar geometry with specific EoS for each
region is given by
\begin{itemize}
\item Inner region $(\mathcal{J}_{1})$\quad $\Rightarrow$\quad $ \mathbf{R}_{1}\geq
r>0$ with $\varrho=-\mathcal{P}$,
\item Thin-shell $(\mathcal{J}_{2})$\quad $\Rightarrow$\quad $\mathbf{R}+\epsilon\geq
r\geq \mathbf{R}_{1}$ with $\mathcal{P}-\varrho=0$,
\item Outer region $(\mathcal{J}_{3})$\quad $\Rightarrow$\quad $r>\mathbf{R}+\epsilon$
with $\varrho=0=\mathcal{P}$.
\end{itemize}

\subsection{Inner Region}

The interior region of charged gravastar satisfies an EoS
$(\mathcal{W}=-1=\mathcal{P}/\varrho)$ that leads the relation
between matter variables as $\mathcal{P}=-\varrho$. The negative
pressure is responsible for outward directed repulsive force to
overcome the pull exerted by thin-shell on the core of spherical
charged gravastar. This particular choice of $\mathcal{W}$ is also
called dark energy EoS. Using this form of EoS in Eq.(\ref{17}), we
get $\varrho=\varrho_{c}$ (where $\varrho_{c}$ is constant) and
hence $\mathcal{P}=-\varrho_{c}$. This relation shows that pressure
and density remain same in the interior of charged gravastar.
Inserting this equation in Eq.(\ref{14}), we acquire the radial
metric function as
\begin{equation}\label{21}
e^{-\xi}=1-\mathcal{H}(r)-\frac{2{\varrho_{c}}{r^{2}}}{3}(4\pi+\zeta{\varrho_{c}})+\frac{C_{1}}{r},
\end{equation}
where $\mathcal{H}(r)$ provides the effect of charge and $C_{1}$
denotes the constant of integration. The singularity-free solution
is obtained at the center of gravastar by setting $C_{1}=0$. Thus we
have
\begin{equation}\label{22}
e^{-\xi}=1-\frac{2{\varrho_{c}}{r^{2}}}{3}(4\pi+\zeta{\varrho_{c}})-
\mathcal{H}(r).
\end{equation}
Using Eqs.(\ref{14}) and (\ref{15}), we can present the correlation
of unknown metric functions as
\begin{equation}\label{23}
e^{\chi}={C_{2}}e^{-\xi},
\end{equation}
where $C_{2}$ corresponds to integration constant. The mass of inner
domain is
\begin{equation}\label{24}
M=\frac{\mathbf{R}^{3}(4\pi\varrho_{c}+\zeta{\varrho_{c}^{2}})}{3}+\int\frac{q^{2}}{r^{2}}dr.
\end{equation}

\subsection{The Intermediate Shell Region}

The non-vacuumed shell of a charged gravastar is filled with an
ultra-relativistic fluid that obeys the EoS $\mathcal{P}-\varrho=0$.
In connection with cold baryonic universe, Zel'dovich introduced the
idea of stiff matter configuration \cite{58aa}. Several researchers
have employed this type of matter distribution and obtained
remarkable results \cite{59aa}-\cite{63aa}. For this region, finding
exact solution to the field equations is difficult. However, in the
limit specified for this domain, i.e., $0<e^{-\xi}\ll1$, such
solutions can be found by taking into account the EoS of this
region. This demonstrates that the smooth matching of inner and
outer domains yields thin-shell presenting the intermediate region
of charged gravastar. By implementing this constraint, the
corresponding field equations can be written as
\begin{eqnarray}\label{25}
\frac{de^{-\xi}}{dr}&=&\frac{2}{r}-\frac{2q^{2}}{r^{3}}-12{\varrho_{0}}^{2}\zeta
r,\\\label{26}
\big(\frac{3}{2r}+\frac{\chi'}{4}\big)\frac{de^{-\xi}}
{dr}&=&\frac{1}{{r}^{2}}-\frac{2q^{2}}{r^{4}}-12{\varrho_{0}}^{2}\zeta.
\end{eqnarray}
Integration of Eq.(\ref{25}) yields
\begin{equation}\label{27}
e^{-\xi}={C_{3}}-6{\varrho_{0}}^{2}\zeta
{r}^{2}+2\ln{r}-2\int\frac{q^{2}}{r^{3}}dr,
\end{equation}
where ${C_{3}}$ indicates another constant of integration. Solving
Eqs.(\ref{25}) and (\ref{26}) to find  $\chi'$ and substituting it
in Eq.(\ref{17}) corresponding to stiff matter configuration
\begin{equation}\label{28}
\mathcal{P}=\varrho=\frac{1}{e^{\chi}}\int \varpi
{e^{\chi}}dr+\mathcal{S}e^{-\chi},
\end{equation}
where $\varpi$ depicts the additional effect of charge and EMSG
gravity and $\mathcal{S}$ is another integration constant. This
relationship implies that fluid distribution at the outer layer of
thin-shell domain is denser than in the inner domain of shell. The
evolution of energy density is illustrated in Figure \textbf{1}. It
can be seen that the energy density is positive across the shell.
\begin{figure}\center
\epsfig{file=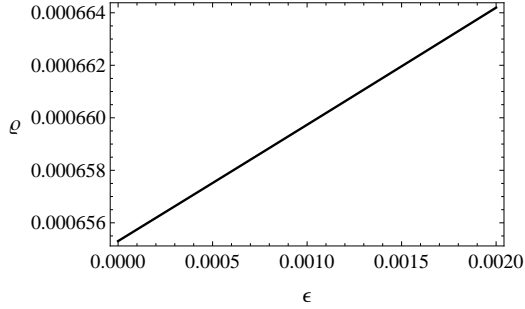,width=0.5\linewidth} \caption{Behavior of
$\varrho$ $(km^{-2})$ versus thickness of thin-shell.}
\end{figure}

\subsection{Outer Region}

We assume that both pressure and density are zero in this region,
i.e., $\varrho=0=\mathcal{P}$. The outer region of spherical charged
gravastar is vacuum and is described by the Reissner-Nordstrom
metric reads as
\begin{equation}\label{29}
(ds)^{2}_{+}=\mathfrak{P}(r) dt^{2}-r^{2}(d\theta^{2}+\sin^{2}\theta
d\phi^{2})-\frac{dr^{2}}{\mathfrak{P}(r)},
\end{equation}
where
$\mathfrak{P}(r)=\frac{r-2\mathcal{M}}{r}+\frac{\mathcal{Q}^{2}}{{r}^{2}}$
and $\mathcal{M}$ corresponds to the whole mass of gravastar. In
order to examine celestial bodies, inner and outer spacetimes must
be perfectly matched. The structure of charged gravastar is mainly
composed of three regions named as $\mathcal{J}_{1}$,
$\mathcal{J}_{2}$ and $\mathcal{J}_{3}$, where $\mathcal{J}_{2}$
acts as a connection between $\mathcal{J}_{1}$ and
$\mathcal{J}_{3}$. The smooth matching of the interior and exterior
geometries is ensured by Israel matching criterion \cite{64aa}. The
continuity of metric coefficients is maintained but it is found that
their derivatives may possess discontinuity at the hypersurface.
Considering Lanczos equations, the stress-energy tensor of matter
surface is computed as
\begin{equation}\label{30}
\mathcal{S}^{\lambda}_{\beta}=\left(\delta^{\lambda}_{\beta}
\mu^{\alpha}_{\alpha}-\mu^{\lambda}_{\beta}\right)\frac{1}{8\pi},\quad
\lambda,\beta=0,2,3,
\end{equation}
where $\lambda, \beta$ denote hypersurface coordinates and
$\mu_{\lambda\beta}=\mathbb{K}^{+}_{\lambda\beta}-\mathbb{K}^{-}_{\lambda\beta}$
implies extrinsic curvature discontinuity. The extrinsic curvature
corresponding to $\mathcal{J}_{1}$ and $\mathcal{J}_{3}$ is defined
as
\begin{equation}\label{31}
\mathbb{K}^{\pm}_{\lambda\beta}=-\left(\frac{\partial^{2}
x^{\tau}}{\partial\Phi^{\lambda}\partial\Phi^{\beta}}+\Gamma^{\tau}_{\mu\nu}
\frac{\partial x^{\mu}\partial x^{\nu}}
{\partial\Phi^{\lambda}\partial\Phi^{\beta}}\right)\Upsilon^{\pm}_{\tau},
\end{equation}
where $\Upsilon^{\pm}_{\tau}$ and $\Phi^{\lambda}$ represent the
unit normals and intrinsic coordinates of the hypersurface,
respectively.

The line element of charged static spherically symmetric geometry is
given as
\begin{equation}\label{32}
ds^{2}=\mathcal{N}(r)dt^{2}-\frac{dr^{2}}{\mathcal{N}(r)}-r^{2}d\theta^{2}
-r^{2}\sin^{2}\theta d\phi^{2}.
\end{equation}
The corresponding unit normals are of the form
\begin{equation}\label{33}
\Upsilon^{\pm}_{\tau}=\pm\left|g^{\mu\nu}\frac{\partial\mathcal{N}
}{\partial x^{\mu}}\frac{\partial\mathcal{N}}{\partial
x^{\nu}}\right|^{-\frac{1} {2}}\frac{\partial\mathcal{N}}{\partial
x^{\tau}},
\end{equation}
\textrm{where} $\Upsilon^{\tau}\Upsilon_{\tau}=1$. For perfect fluid
matter configuration, we obtain
$\mathbb{S}_{\lambda\beta}=$diag$(\Omega,$ $-\Psi,-\Psi)$. Hence,
the surface density turns out to be
\begin{equation}\label{34}
\Omega=-\frac{1}{4\pi\mathbf{R}}\Big[\sqrt{1-\frac{2\mathcal{M}}{\mathbf{R}}+
\frac{\mathcal{Q}^{2}}{\mathbf{R}^{2}}}
-\sqrt{1-\frac{(4\pi+\zeta{\varrho_{c}}){2{\varrho_{c}}{\mathbf{R}^{2}}}}{3}+\mathcal{H}(r)}\Big].
\end{equation}
Using $\mathcal{J}_{1}$ and $\mathcal{J}_{3}$ as domains of charged
gravastar, we obtain respective expression of surface pressure as
\begin{equation}\label{35}
\Psi=\frac{1}{8\pi\mathbf{R}}\Big[\frac{{1-\frac{\mathcal{M}}{\mathbf{R}}}}
{\sqrt{1-\frac{2\mathcal{M}}{\mathbf{R}}+\frac{\mathcal{Q}^{2}}{\mathbf{R}^{2}}}}-
\frac{1-\frac{(4\pi+\zeta{\varrho_{c}}){4{\varrho_{c}}{\mathbf{R}^{2}}}}
{3}+\mathcal{H}(r)+\mathcal{H'}(r)}{\sqrt{1-\frac{(8\pi+2\zeta{\varrho_{c}})
{{\varrho_{c}}{\mathbf{R}^{2}}}}{3}+\mathcal{H}(r)}}\Big].
\end{equation}
The mass inside gravastar is calculated through Eq.(\ref{34}) as
follows
\begin{equation}\label{36}
{m}_{\textit{shell}}=4\pi\mathbf{R}^{2}\Omega=\mathbf{R}
\Big[{\sqrt{1-\frac{(4\pi+\zeta{\varrho_{c}})
{2{\varrho_{c}}{\mathbf{R}^{2}}}}{3}+\mathcal{H}(r)}}
-\sqrt{1-\frac{2\mathcal{M}}{\mathbf{R}}+\frac{\mathcal{Q}^{2}}{\mathbf{R}^{2}}}\Big].
\end{equation}
Finally, the total mass of charged gravastar is
\begin{eqnarray}\nonumber
\mathcal{M}&=&{m}_{\textit{shell}}{\sqrt{1-\frac{(4\pi+\zeta{\varrho_{c}})
{2{\varrho_{c}}{\mathbf{R}^{2}}+\mathcal{H}(r)}}{3}}}+\frac{(4\pi+\zeta{\varrho_{c}})
{2{\varrho_{c}}{\mathbf{R}^{3}+\mathcal{H}(r)}}}{3}\\\label{37}
&-&\frac{{m}_{\textit{shell}}^{2}}{2\mathbf{R}}
-\frac{\mathcal{Q}^{2}}{\mathbf{R}^{2}}.
\end{eqnarray}

\section{Attributes of Charged Gravastars}

This section deals with the study of characteristics of charged
gravastar, i.e., the EoS parameter, entropy, proper length and
thin-shell energy through graphical analysis. We have analyzed these
features of thin-shell for negative as well positive values of the
model parameter for three different values of charge. This
discussion will show viability of the spherical charged gravastar
structure in the context of EMSG.

\subsection{The EoS Parameter}

The EoS for shell of charged gravastars at $r=\mathbf{R}$ is
\begin{equation}\label{ww}
\mathcal{W}=\frac{\Psi}{\Omega}.
\end{equation}
Replacing the values of $\Psi$ and $\Omega$ in Eq.(\ref{ww}), we
have
\begin{equation}\label{jj}
\mathcal{W}(\mathbf{R})=\frac{\frac{1}{8\pi\mathbf{R}}\Big[\frac{{1-\frac{\mathcal{M}}{\mathbf{R}}}}
{\sqrt{1-\frac{2\mathcal{M}}{\mathbf{R}}+\frac{\mathcal{Q}^{2}}{\mathbf{R}^{2}}}}-
\frac{1-\frac{(4\pi+\zeta{\varrho_{c}}){4{\varrho_{c}}{\mathbf{R}^{2}}}}
{3}+\mathcal{H}(r)+\mathcal{H'}(r)}{\sqrt{1-\frac{(8\pi+2\zeta{\varrho_{c}})
{{\varrho_{c}}{\mathbf{R}^{2}}}}{3}+\mathcal{H}(r)}}\Big]}{-\frac{1}{4\pi\mathbf{R}}
\Big[\sqrt{1-\frac{2\mathcal{M}}{\mathbf{R}}+
\frac{\mathcal{Q}^{2}}{\mathbf{R}^{2}}}
-\sqrt{1-\frac{(4\pi+\zeta{\varrho_{c}}){2{\varrho_{c}}{\mathbf{R}^{2}}}}{3}+\mathcal{H}(r)}\Big]}.
\end{equation}
We impose some additional constraints on $\mathcal{W}(\mathbf{R})$,
i.e.,
$\frac{2\mathcal{M}}{\mathbf{R}}+\frac{\mathcal{Q}^{2}}{\mathbf{R}^{2}}<1$
or $2\mathcal{M}<\mathbf{R}$. The relationship between
$\mathcal{M}$, $\mathcal{Q}$ and $\mathbf{R}$ is obtained as
$\mathcal{Q}>\sqrt{2\mathcal{M}\mathbf{R}-\mathbf{R}^{2}}$ and
$2\mathcal{M}<\mathbf{R}$. Taking positive values of surface density
as well as pressure results in positive value of $\mathcal{W}$. We
attain $\mathcal{W}(\mathbf{R})\approx1$ for large values of
$\mathbf{R}$. The structure of gravastar is similar to compact
objects for appropriately bigger values of  $\mathbf{R}$. Moreover,
by inserting specific value of $\mathbf{R}$ in Eq.(\ref{35}) may
result $\Psi=0$, which presents dust shell.

\subsection{Entropy of Thin-Shell}
\begin{figure}\center
\epsfig{file=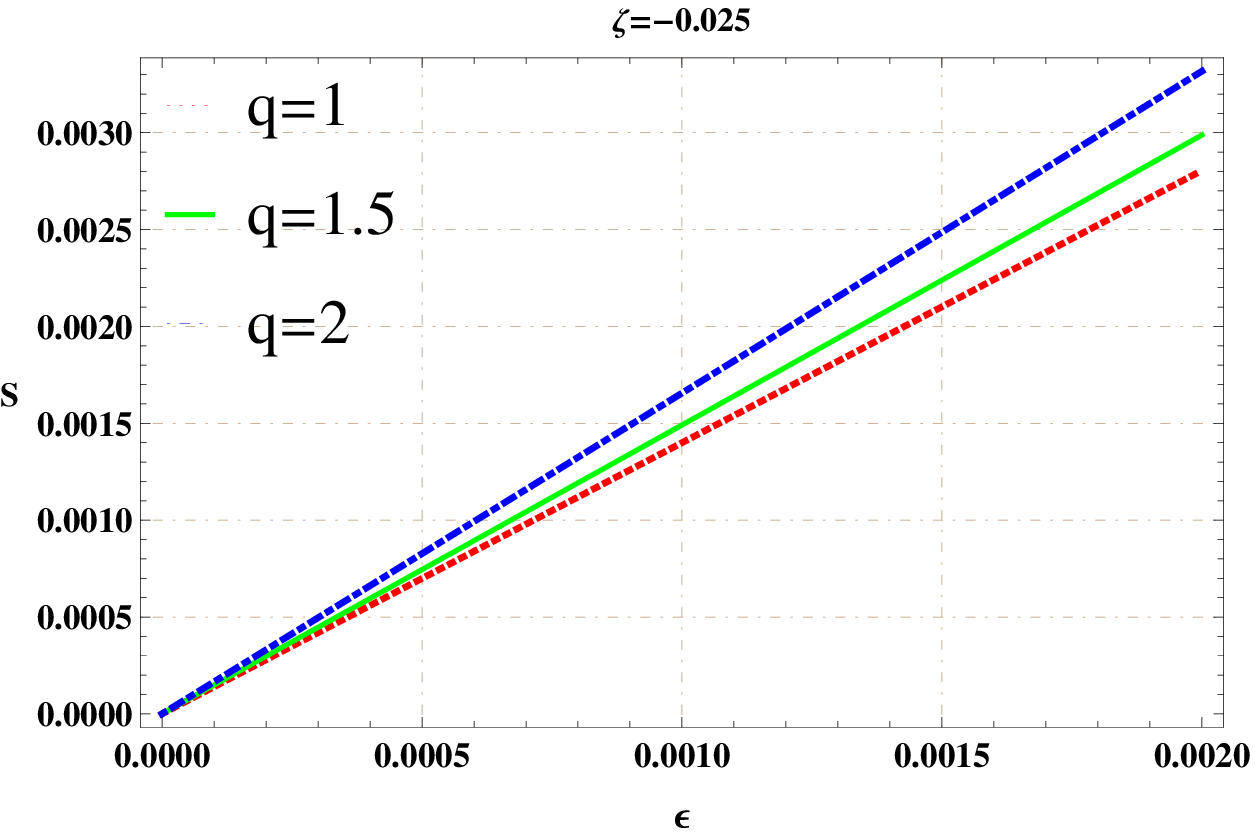,width=.5\linewidth}\epsfig{file=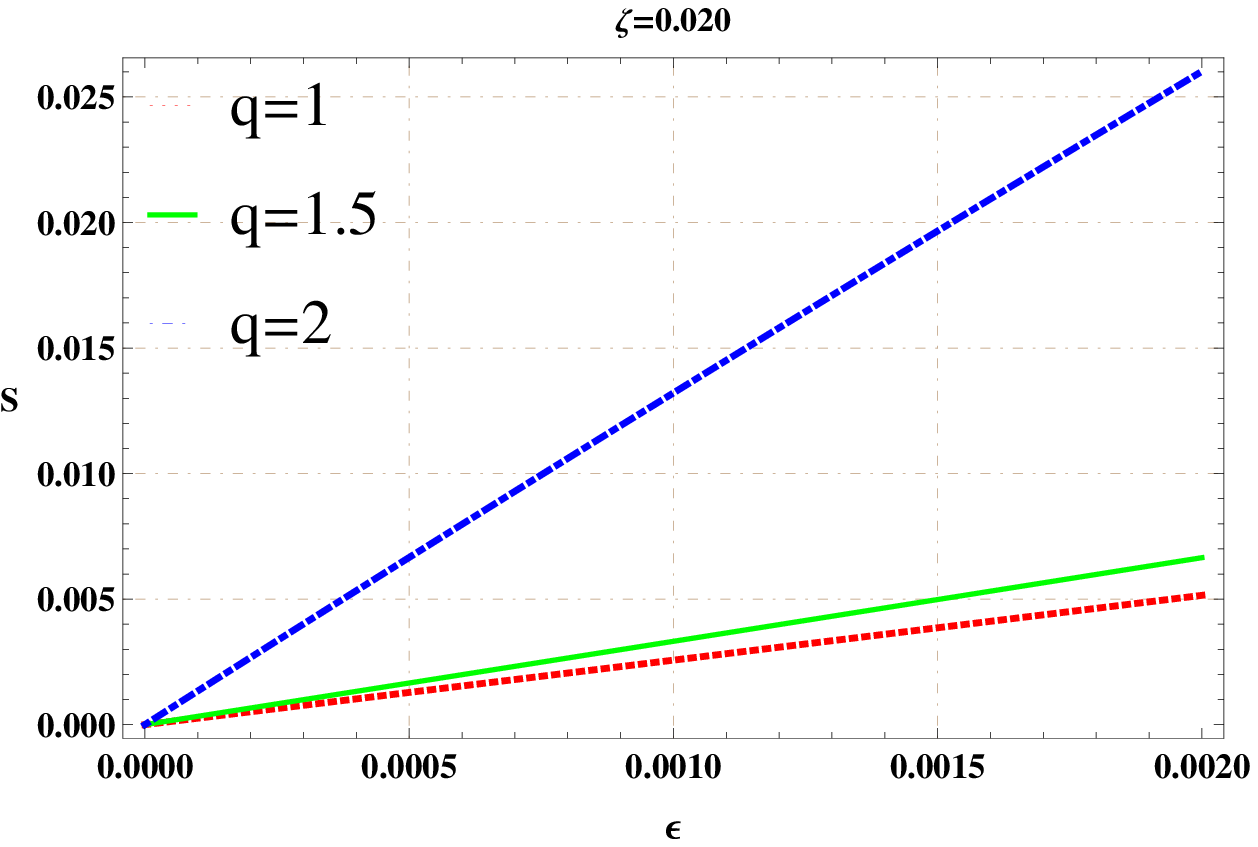,width=.49\linewidth}
\caption{Entropy of the shell with thickness (km) corresponding to
different values of charge.}
\end{figure}

Mazur and Mottola \cite{13a} investigated the inner region and
concluded that entropy density is zero which represents reliability
of condensate phase. In intermediate region, the entropy of charged
gravastar is calculated as
\begin{equation}\label{ea}
S=\int^{\mathbf{R}_{2}}_{\mathbf{R}_{1}}4\sqrt{e^{\xi(r)}}\pi
r^{2}\mathbb{V}(r)dr,
\end{equation}
where $\mathbb{V}(r)$ shows the entropy density given as
\begin{equation}\label{}
\mathbb{V}(r)=\frac{\curlyvee^{2}K^{2}_{\mathcal{B}}\mathbb{T}(r)}{4\pi
\hslash^{2}}=K_{\mathcal{B}}
\left({\frac{\mathcal{P}}{2\pi}}\right)^{\frac{1}{2}}\frac{\curlyvee}{\hslash},
\end{equation}
$\curlyvee$ is a dimensionless constant. The Planckian units are
taken into account $K_{\mathcal{B}}=\hslash=1$ so that Eq.(\ref{ea})
becomes
\begin{equation}\label{}
S=\curlyvee({8\pi})^{\frac{1}{2}}\int^{\mathbf{R}_{2}}_{\mathbf{R}_{1}}\frac
{({C_{3}}-6{\varrho_{0}}^{2}\zeta
{r}^{2}+2\ln{r}-2\int\frac{q^{2}}{r^{3}}dr)^{\frac{-1}{2}}}{(\frac{1}{e^{\chi}}\int
\varpi {e^{\chi}}dr+\mathcal{S}e^{-\chi})^{\frac{-1}{2}}}{r}^{2}dr.
\end{equation}
It is too difficult to obtain an analytical solution to this
equation, therefore, we solve it numerically. Figure \textbf{2}
depicts the disorderness of gravastar and shows that the entropy of
shell increases with the increment in thickness. The entropy is
proportional to the small thickness of thin-shell. It has maximum
value at the outer surface of thin-shell and zero thickness yields
the value of entropy to be zero. The disorderness increases with the
increasing amount of charge, thus allowing the formation of a more
stable structure. The model parameter affects the entropy such that
its positive value provides higher entropy as compared to its
negative value.

\subsection{Length of Charged Shell}
\begin{figure}\center
\epsfig{file=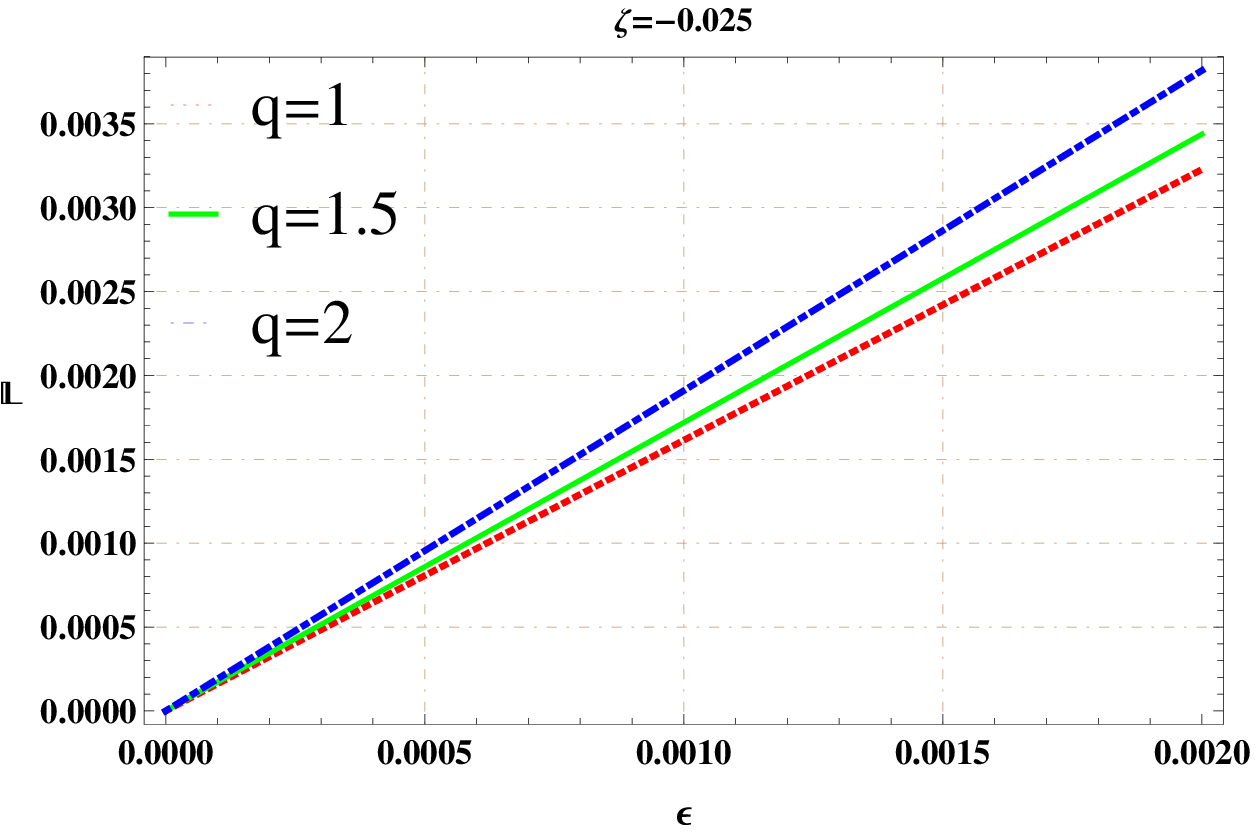,width=.5\linewidth}\epsfig{file=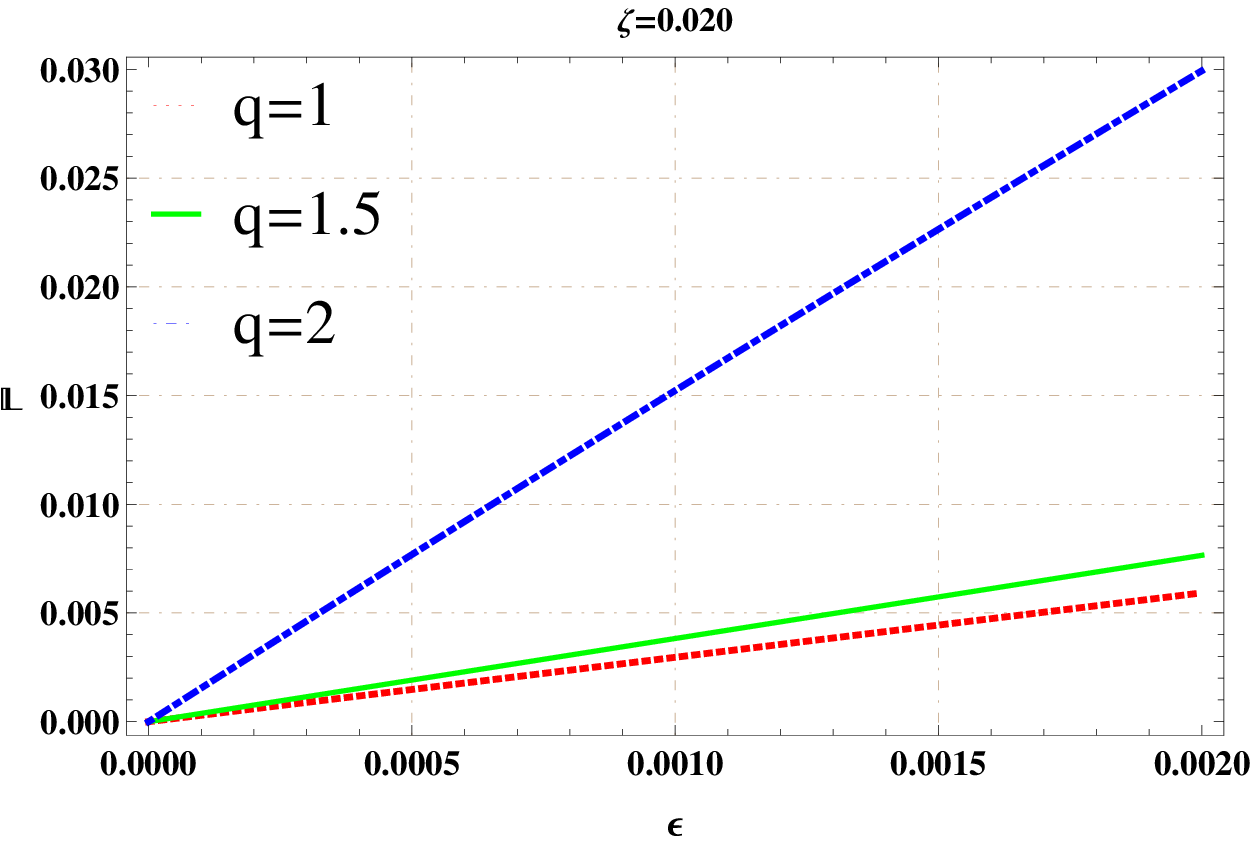,width=.5\linewidth}
\caption{Proper length (km) of the charged shell against thickness
(km).}
\end{figure}

The proper length of shell region is calculated from
$\mathbf{R}_{1}$ to $\mathbf{R}_{2}=\mathbf{R}+\epsilon$, where
$\epsilon\ll1$ indicates negligible change. The mathematical formula
of length of intermediate-shell is given by
\begin{equation}\label{}
\mathbb{L}=\int^{\mathbf{R}_{2}}_{\mathbf{R}_{1}}\sqrt{e^{\xi(r)}}dr
=\int^{\mathbf{R}_{2}}_{\mathbf{R}_{1}}\frac{dr}{\sqrt{{C_{3}}-6{\varrho_{0}}^{2}\zeta
{r}^{2}+2\ln{r}-2\int\frac{q^{2}}{r^{3}}dr}}.
\end{equation}
We obtain the numerical solution of this equation and its plot is
shown in Figure \textbf{3}. The left panel of Figure \textbf{3}
shows a continuous increment in the length of thin-shell by
increasing the value of charge with $\zeta=-0.025$. However,
$\zeta=0.020$ leads to the rapid increase in length of the shell for
similar values of charge, presented in the right panel of Figure
\textbf{3}.

\subsection{Thin-Shell Energy}

In the core of charged gravastar, the presence of dark energy is
confirmed due to negative pressure which is responsible for
repulsive force. The mathematical representation of the energy is
\begin{equation}\label{ey}
\varepsilon=\int^{\mathbf{R}_{2}}_{\mathbf{R}_{1}}4\pi\varrho
r^{2}dr.
\end{equation}
Plugging the value of surface density in Eq.(\ref{ey} yields
\begin{equation}\label{}
\varepsilon=\int^{\mathbf{R}_{2}}_{\mathbf{R}_{1}}4\pi
{r}^{2}\left(\frac{1}{e^{\chi}}\int \varpi
{e^{\chi}}dr+\mathcal{S}e^{-\chi}\right)dr.
\end{equation}
This increases by enhancing its thickness as shown in Figure
\textbf{4}. The energy plots corresponding to both values of $\zeta$
show linear profile with the increment of charge.
\begin{figure}\center
\epsfig{file=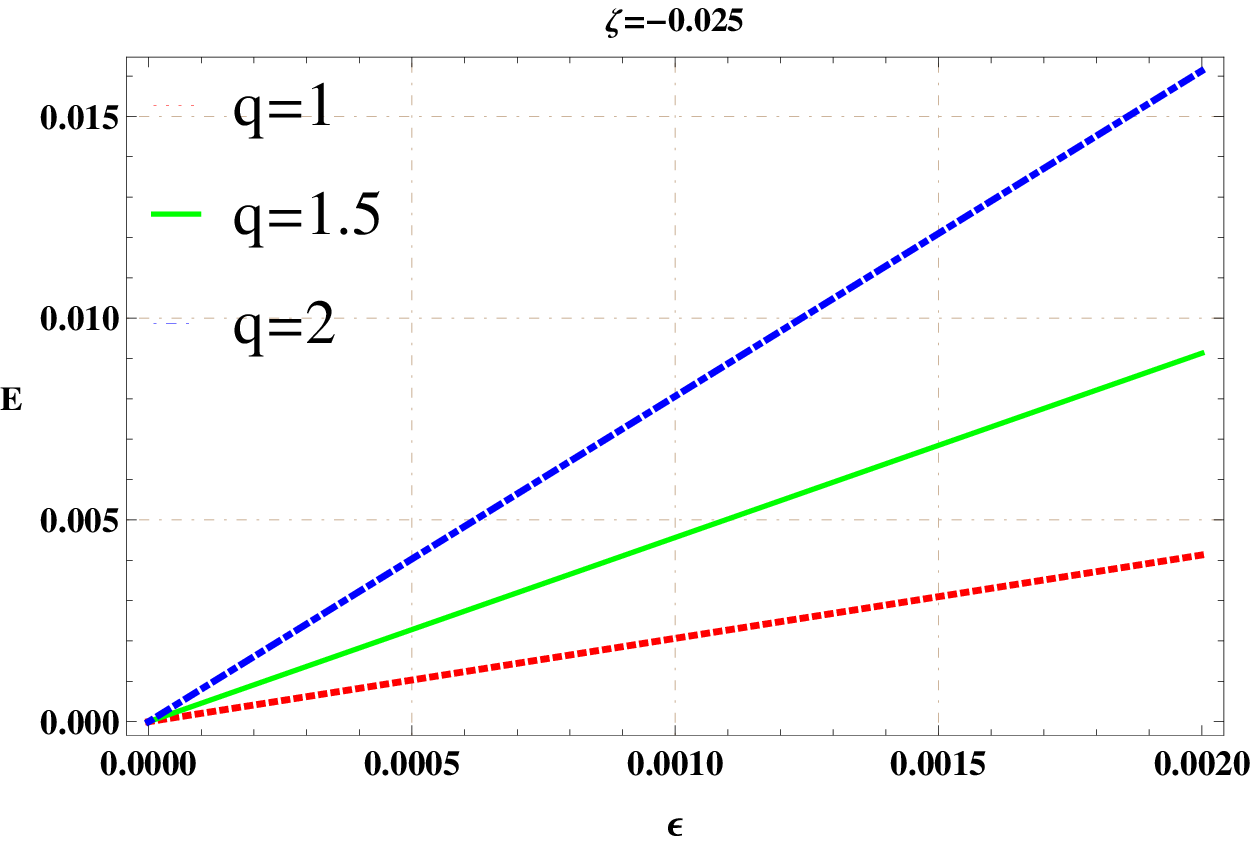,width=.5\linewidth}\epsfig{file=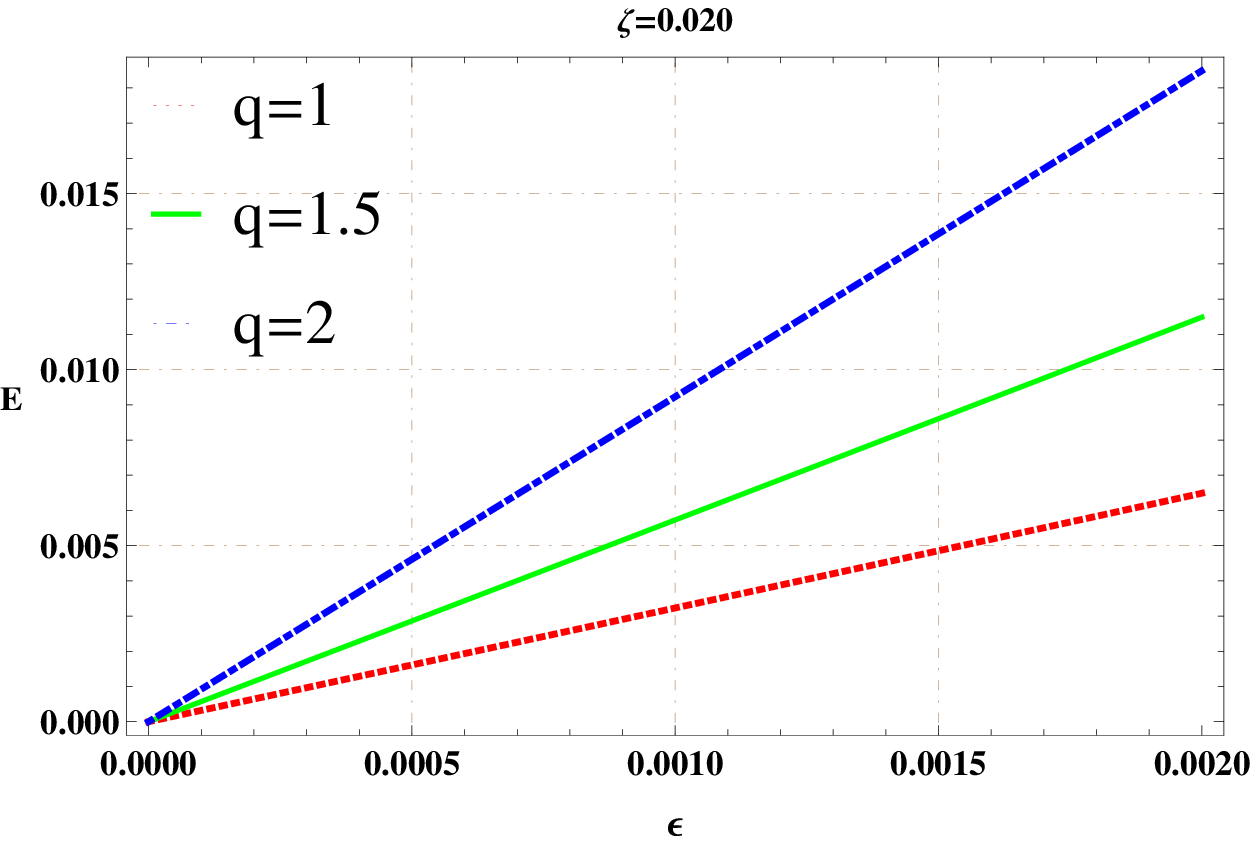,width=.5\linewidth}\caption{Plots
of energy (km) of charged thin-shell versus thickness of the shell
(km).}
\end{figure}

\section{Conclusions}

In this manuscript, we investigate the structure of charged
gravastar in the background of EMSG. We have considered charged
sphere in the interior with EoS $\varrho=-\mathcal{P}$ and
Reissner-Nordstrom black hole in the exterior region with vacuum
($\varrho=\mathcal{P}=0$). The interior region is surrounded by a
thin-shell with EoS $\varrho=\mathcal{P}$. In the intermediate
shell,  non-repulsive force counterbalances the pull exerted by the
charged sphere and assures the formation of singularity-free compact
object. It is worth mentioning that solutions corresponding to dark
energy EoS show connection of charged gravastars with dark stars.

It is worth here that solutions obtained through dark energy EoS
indicate that charged gravastars are linked with dark stars.

The main results are summarized as follows.
\begin{itemize}
\item The increasing density depicts that the shell's
external boundary is denser than the internal boundary (Figure
\textbf{1}).
\item The real value of the EoS parameter is obtained when
$\frac{2\mathcal{M}}{\mathbf{R}}+\frac{\mathcal{Q}^{2}}{\mathbf{R}^{2}}<1$.
This is a necessary condition for the stable and physically viable
gravastar structure.
\item The entropy of charged thin-shell is proportional to its thickness,
thus greater thickness means higher entropy (Figure \textbf{2}). The
increment of electromagnetic strength increases its entropy as well.
\item
The proper length of the charged thin-shell indicates increasing
behavior with the thickness of the shell (Figure \textbf{3}). The
gravastar length goes on increasing with a higher value of the
charge.
\item
The energy of the inner boundary is less than that of the outer
domain. The thickness of the shell is shown to be linearly related
to the energy of the shell (Figure \textbf{4}).
\end{itemize}

It is worthwhile to mention here that the length, energy and entropy
of a gravastar increase with respect to thickness of the shell as
compared to the uncharged case \cite{w7}. Our analysis follows a
consistent increasing trend of the physical properties in the
presence of charge, presented in GR as well as other modified
theories of gravity \cite{7dddd,cg8,w3,w4}. The matching of the
charged inner sphere and Reissner-Nordstrom metric produces
gravitational mass which behaves as an electromagnetic mass model.
The charge provides an outward-directed force, therefore, an
additional repulsive force helps the gravastar from collapsing into
a singularity. Thus, the presence of charge generates a more stable
structure as compared to the uncharged model \cite{w7}. Furthermore,
we have found that the physical attributes of gravastar have greater
values in EMSG as compared to GR and other modified theories of
gravity \cite{7dddd,cg8,w3,w4}. We conclude that
$f(\mathfrak{R},\mathcal{T}^{2})$ gravity successfully discusses
charged gravastar model.

\end{document}